\documentstyle[12pt,cite,epsfig]{article}

\renewcommand{\today}{11th March  1999}

\newcommand{\nc}{\newcommand}

\nc{\be}{\begin{equation}}
\nc{\ee}{\end{equation}}
\nc{\bea}{\begin{eqnarray}}
\nc{\eea}{\end{eqnarray}}
\nc{\beas}{\begin{eqnarray*}}
\nc{\eeas}{\end{eqnarray*}}
\nc{\noi}{\noindent}
\nc{\sD}{\not \! \! D}
\nc{\s}[1]{\not \! #1}
\nc{\non}{\nonumber}
\nc{\bb}{\bibitem}
\nc{\lf}{\left}
\nc{\ri}{\right}
\nc{\mb}[1]{\makebox[#1]{}}
\nc{\pa}{\partial}
\nc{\sA}{\not \! \! A}
\nc{\newsec}[1]{\section{#1}\mb{0.5cm}}
\nc{\h}{\frac{1}{2}}
\nc{\ra}{\rightarrow}
\nc{\la}{\leftarrow}
\nc{\ep}{$e^+e^-\ra\pi^+\pi^-\;$}
\nc{\emuon}{$e^+e^-\ra\mu^+\mu^-\;$}
\nc{\epp}{$e^+e^-\ra\pi^+\pi^0\pi^-\;$}
\nc{\elec}{$e^+e^-\ra\gamma^*\ra e^+e^-\;$}
\def\mathunderaccent#1{\let\theaccent#1\mathpalette\putaccentunder}
\def\putaccentunder#1#2{\oalign{$#1#2$\crcr\hidewidth
\vbox to.2ex{\hbox{$#1\theaccent{}$}\vss}\hidewidth}}

\nc{\ti}{\mathunderaccent\tilde}
\nc{\M}{{\cal M}}
\nc{\rw}{$\rho\!-\!\omega\;$}
\def\hhht{\rule[ 0.mm]{0.mm}{6.mm}}

\def\hhhb{\rule[-3.mm]{0.mm}{9.mm}}
\def\hhhc{\rule[-3.mm]{0.mm}{3.mm}}
\def\hhhd{\rule[-3.mm]{0.mm}{1.mm}}
\def\hhhu{\rule[-3.mm]{0.mm}{12.mm}}

\begin{document}
\thispagestyle{empty}
\begin{flushright}
                                       hep-ph/9906372
\end{flushright}                                         

\begin{center}
\pagenumbering{arabic}
{\large{\bf A VMD Based, Nonet and SU(3) Symmetry Broken \\[0.5cm]
Model For Radiative Decays of Light Mesons}}\\
\vspace{1.0 cm}
talk presented at\\[0.5cm]
{\bf The International Workshop ``$e^+e^-$ Collisions from 
$\phi$ to $J/\psi$''\\1-5 Mar 1999,  Novosibirsk, Russia }\\[0.5cm]
by\\[0.5cm]
{\bf M.~Benayoun}\\[0.5cm]
in name of\\[0.7cm]
M.~Benayoun$^{a}$, L. DelBuono$^{a}$,
        S. Eidelman$^{a,b}$, V. N. Ivanchenko$^{a,b}$, H.B. O'Connell$^{c}$
\vspace{1.2 cm}

$^{a}$LPNHE des Universit\'es Paris VI et VII--IN2P3, Paris,
         France\\
         $^{b}$Budker Institute of Nuclear Physics, Novosibirsk 630090, 
         Russia \\
         $^{c}$Stanford Linear Accelerator Center, Stanford University,
         Stanford CA 94309, USA
\vspace{1.0 cm}
\date{\today}
\begin{abstract}
We present a VMD based model aiming to describe all radiative decays 
of light mesons.
We show that the SU(3) breaking mechanism proposed by Bando, Kugo 
and Yamawaki (BKY),
supplemented  by nonet symmetry breaking in the pseudoscalar sector
are sufficient to provide a nice description of all data, except 
the $K^{*\pm}$ radiative width.
It is also shown that nonet symmetry breaking has effects which cannot be 
disantangled from those produced by coupling of glue to the $\eta'$ meson. 
Coupling of glue to $\eta$ is not found to be required by the data. 
 Asssuming the $K^{*\pm}$ radiative width is indeed at
its presently accepted value necessitates to supplement the BKY
breaking in a way which finally preserves an equivalence statement between
the VMD approach to radiative decays and the Wess--Zumino--Witten Lagrangian. 
\end{abstract}
\end{center}

\newpage

\pagenumbering{arabic}
\section{Introduction.}

\indent \indent
It is a long standing problem to define a framework in
which all radiative decays of light flavor mesons can be
accurately accounted for.  A few kinds of different models have been 
proposed so far. The most popular modelling is in terms of magnetic 
moments of quarks \cite{dolinsky,volodia}. 
Another traditional approach is to use SU(3) relations 
among coupling constants \cite{odonnel}. This yields
reasonable descriptions of radiative decays \cite{ben2},
though the success is not complete. 

The O'Donnell model
assumes exact SU(3) flavor symmetry, while
nonet (or U(3) flavor) symmetry is explicitly broken. 
As it follows from a quite general conceptual framework,
this model is widely independent of detailed dynamical properties
and assumptions. 
This model covers  all couplings like $PV\gamma$ but lacks to describe 
$P \gamma \gamma$ decays which remain unrelated.

Recently, several models of other kinds have been proposed \cite{BGP,ball}, 
motivated by effective Lagrangian approaches to the interactions of vector mesons 
\cite{HLS,FKTUY}, with various kinds of SU(3) breaking schemes \cite{BGP,BKY,ball,heath}.
However, breaking of nonet symmetry in phenomenological models
has got little attention \cite{odonnel,ben2}.

The study of radiative decays of light flavor mesons is also 
connected with the long standing problem of $\eta/\eta'$ mixing 
\cite{ben2,DHL,GILMAN} and to its possible association with a glue 
component inside light mesons \cite{ball,veneziano}. 
Recent developments advocate a more complicated 
$\eta/\eta'$ mixing scheme \cite{leutw,leutwb}. 

As  effects of SU(3) symmetry breaking are clearly observed in the
data on radiative decays of light  mesons \cite{ben2,ball}, they have surely 
to be accounted for. We do it following the BKY breaking mechanism \cite{BKY,heath}.
 In this way, by means of the FKTUY Lagrangian and of the BKY breaking scheme, 
we can construct a Lagrangian formulation of the O'Donnell model 
and extend it to the case where the SU(3) flavor symmetry is broken.
This, additionally, provides an algebraic connection between
$VP\gamma$ and $P\gamma \gamma$ coupling constants.
  
What is presented here is an account of a
work \cite{rad} mainly devoted to a study of radiative of  light flavor mesons 
within the general VMD framework of the hidden local symmetry model \cite{HLS}
(hereafter referred to as HLS) and its anomalous sector \cite{FKTUY}
(hereafter referred to as FKTUY). All details and previous references
can be found there. The perspective of this talk is however
somewhat different, due to recent developments presented in Ref.\cite{chpt}.

\begin{figure}[htb]
  \centering{\
     \epsfig{angle=0,figure=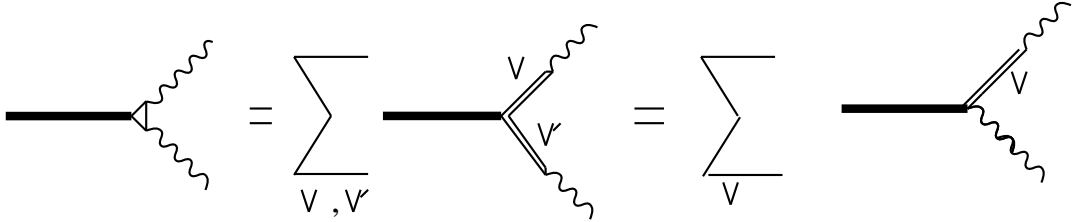,width=0.9\linewidth}
                    }
\parbox{130mm}{\caption
{Graphical representation of the relation among various kind
of coupling constants. $V$ and $V'$ stand for the lowest lying
vector mesons ($\rho^0$, $\omega$, $\phi$)~; the internal 
vector meson lines are propagators at $s=0$ and are approximated
by the corresponding tabulated \cite{PDG98} masses squared.
}
\label{graph}}
\end{figure}

The HLS model is an expression of the Vector Meson Dominance
(VMD) assumption~; it thus gives a way to relate the 
radiative decay modes $VP\gamma$ to each other and to the
$P\gamma \gamma$ decays  for light mesons, by giving a precise
meaning to the equations shown in Fig. \ref{graph}.

One can try to estimate naively this relation
using other information
collected in the Review of Particle Properties \cite{PDG98}. 

\begin{center} 
\begin{tabular}{|| c  | c  | c |c | c |c ||}
\hline
\hline
\hhhc Mode & VMD prediction  &  PDG\\
\hline
\hline
$\pi^0 \rightarrow \gamma \gamma$\hhhu [eV]&
 $ 10.73 \pm 1.20$ & $7.74 \pm 0.50$ \\
\hline
$\eta \rightarrow \gamma \gamma$\hhhu [keV]&
$0.62 \pm 0.18$ & $0.46 \pm 0.04$ \\
\hline
$\eta' \rightarrow \gamma \gamma$\hhhu[keV] &
$5.10 \pm 0.76$ & $4.27 \pm 0.19$\\
\hline
\end{tabular}
 
\parbox[t]{16.0cm}{ \hhht
      {\bf Table 1} :  Partial decay widths of the pseudoscalar
      mesons, as reconstructed from VMD, using the $VP\gamma$
      measured couplings, and their direct accepted measurements
      \cite{PDG98}. 
}  
\end{center}

The results given in Table 1 are quite impressive~;  
this is indeed not a fit, but mere algebra. Thus, all
systematics can pill up and, moreover,  the meson masses used in order to 
estimate the propagators at $s=0$ are  simply the (Breit--Wigner)
accepted masses \cite{PDG98}~; this is surely a very crude assumption, 
at least for the $\rho^0$ meson.  However, this exercise teaches us that 
the central hint of the Vector Meson Dominance assumption  is sharply 
grounded and this motivates to try going beyond as much as possible. 

A further comment is of relevance concerning the VMD prediction
for the $\pi^0$ decay width. Actually, the two--photon width of 
the $\pi^0$ can be computed, as sketched in Fig. 1, from two different
ways since the basic $VVP$ diagram is $\pi^0 \rho^0 \omega^I$,
where $\omega^I$ is the ideal (purely non--strange) combination
of the (physical) $\omega$ and $\phi$ fields. A first estimate
is thus obtained from the coupling $\pi^0 \rho^0 \gamma$
(here the hidden vector meson line is surely $\omega^I$)
and is $12.79 \pm 2.59$ eV~; a second estimate is obtained
from using instead the couplings $\pi^0 \omega \gamma$
and $\pi^0 \phi \gamma$ (here the hidden vector meson line is 
surely $\rho^0$ for both) and is $8.86 \pm 0.29$ eV. The qualitative
difference of both estimates reflects problems with the
$\rho$ mass definition which will not be examined here.
What is given in Table 1 is simply the mean value of
both estimates.

\section{An Exact SU(3) Symmetry Framework}
\label{exactsymm}

\indent \indent The formalism which describes the decays 
$V \rightarrow P ~\gamma$
and $ P \rightarrow V \gamma$ within an exact  SU(3) flavor symmetry framework 
has been given by P. O'Donnell in Ref.~\cite{odonnel}. 
The corresponding decay amplitudes can be quite generally written as
\begin{equation}
T=g_{VP\gamma} \epsilon_{\mu \nu \rho \sigma}k^{\mu} q^{\nu} 
\varepsilon^{\rho}(V)\varepsilon^{\sigma}(\gamma) 
\label{model1} 
\end{equation}

\noindent
using obvious notations. This expression can be found by relying
entirely on gauge invariance and does not require the help of any
specific Lagrangian.

Using SU(3) symmetry, the coupling strengths $g_{VP\gamma}$
between physical vector and pseudoscalar mesons in radiative decays
are expressed in terms of two angles ($\theta_V$ and $\theta_{P}$)
which describe the mixtures of singlet and octet components,
and of three coupling constants ($g_{V_8P_8\gamma}$, $g_{V_0P_8\gamma}$
and $g_{V_8P_0\gamma}$); as the photon behaves like an SU(3)
octet, this cancels out the possible coupling $g_{V_0P_0\gamma}$.
We do not reproduce here the expressions for the  $g_{VP\gamma}$
in terms of the elementary couplings $g_{V_iP_j\gamma}$ and the mixing
angles~; they can be found in Ref.~\cite{odonnel} and 
in Appendix A7 of Ref.~\cite{ben2}, where a misprint has been corrected.

One generally uses the representation of the matrices for vector mesons
($V$) and pseudoscalar mesons ($P$) in the $\{u$, $d$, $s\}$ basis.
Their expressions are quite classical and can be found, for instance, in 
Refs. \cite{heath,rad}. It is usual to express the relevant matrix
elements of $V$ in terms of  ideally mixed states ($\omega^I$, $\phi^I$), while
it is as traditional to express the isoscalar mesons fields in $P$
in terms of the conventional octet and singlet components
($\pi_8$, $\eta_0$). These are not the physical states ($\omega$, $\phi$, 
 $\eta$, $\eta'$) which are generated from 
ideally mixed states by means of rotations. The rotation angles from singlet and octet 
states to the physically observed mesons are traditionally named $\theta_V$ and $\theta_P$.
These well known relations can be found in Refs. \cite{odonnel,heath,rad,PDG98}.

With these definitions for the field matrices, the effective FKTUY Lagrangian
which describes the anomalous sector of the HLS model is \cite{FKTUY}

\begin{equation}
{\cal L}=- \frac{3 g^2}{4 \pi^2 f_{\pi}} 
\epsilon^{\mu \nu \rho \sigma} {\rm Tr}[\pa_\mu V_\nu\pa_\rho V_\sigma P].
\label{fktuy0}
\end{equation}

The universal vector meson coupling $g$ 
is tightly related to the coupling of the $\rho$ meson to a pion pair and
$f_{\pi}=92.41$ MeV is the usual pion decay constant. 
The partial widths for  all $VP\gamma$ and $P \gamma \gamma$ modes
are derived herefrom, using also  the $V \gamma$ transition amplitudes
and the expressions for the vector meson masses
given in the standard (non--anomalous) HLS Lagrangian \cite{HLS,rad}.
Actually, the coefficient in Rel. 
(\ref{fktuy0}) is fixed \cite{FKTUY} in order that this Lagrangian  leads to
the usual expression for the amplitude of $\pi^0 \rightarrow \gamma \gamma$.

At this point, it should be emphasized that exact SU(3) symmetry is 
not in conflict with releasing the condition of nonet symmetry
(which corresponds to the stronger U(3) symmetry)
usually stated in effective Lagrangian models for both the vector and 
pseudoscalar meson sectors \cite{HLS,FKTUY}. 
The field matrices $V$ and $P$ can be written $V=V_8+V_0$
and $P=P_8+P_0$, in order to exhibit  their (matrix) octet
and singlet mixtures. It can be checked, that the O'Donnell
model \cite{odonnel,rad} can be generated by simply replacing in Rel. 
(\ref{fktuy0}) the (nonet symmetric) vector and pseudoscalar
field matrices by

\begin{equation}
\left\{
\begin{array}{ll}
P=P_8+xP_0 \\[0.5cm]
V=V_8+yV_0 
\end{array}
\right.
\label{fktuy1}
\end{equation}

Referring to \cite{odonnel,rad}, the basic coupling constants of the model are

\begin{equation}
\left\{
\begin{array}{lll}
g_{V_8P_8\gamma}= & ~~G=& \displaystyle -\frac{3eg}{8 \pi^2 f_{\pi}}\\[0.5cm]
g_{V_0P_8\gamma}= & yG &  \\[0.5cm]
g_{V_8P_0\gamma}= & xG &  
\end{array}
\right.
\label{rad0mod}
\end{equation}

So, one can choose $x$, $y$ and $G$
in addition to the mixing angles as free parameters to be determined from fit to
data. This identification is already interesting, as it relates 
the main coupling constant $G$ in the O'Donnell model to more usual
quantities ($g$, $f_{\pi}$). Finally, $x$ and $y$ are the deviations
from nonet symmetry in respectively the pseudoscalar and vector sectors. 

It was phenomenologically checked{\footnote{The quoted deviations of $x$ and
$y$ from unity have been confirmed by the present analysis.
For instance, releasing $y$ in the present U(3), SU(3) broken model
leads to $y=0.996 \pm 0.033$, quite consistent indeed with $y=1$
and thus with nonet symmetry.}} 
\cite{ben2}, that radiative decays
of light mesons are consistent with nonet symmetry in the vector sector~;
thus, data accommodate $y=1$ quite naturally. However, the same analysis
concluded to a small but significant departure (slightly more than $4\sigma$) 
from nonet symmetry in the pseudoscalar sector, which corresponds to  
$x \simeq0.90$. From a physics point of view,
the resulting picture was close to be acceptable, as only two decay
modes, $K^{*0} \rightarrow K^0 \gamma$ and $\phi \rightarrow \eta \gamma$,  
were not satisfactorily accounted for (see in Ref. \cite{ben2} the ``internal fit'' entry
of Table 8). Qualitatively, the former disagreement could be due to SU(2) symmetry breaking 
because of the $K^{*0}$ quark content{\footnote{However, a factor
of 1.5 at the rate level, {\it i.e.} a factor 1.25 at the level
of coupling constants, could look somehow beyond expectable SU(2) breaking effects.}}.
However, the disagreement about the later mode (more than a factor of 2) is clearly a 
signal of unaccounted for  SU(3) breaking effects, since the branching fraction for 
$\phi \rightarrow \eta \gamma$ has been recently confirmed twice \cite{PDG98}.

Therefore the O'Donnell model \cite{odonnel,ben2} is already close enough
to observations that one may conclude that nonet symmetry breaking is a working concept 
and that only some amount of SU(3) breaking is needed in order to achieve a quite
consistent description of radiative decays. This is the main purpose
of the work presented here. It should be re--emphasized, however, that the original
model of O'Donnell did not relate $VP\gamma$ and $P\gamma\gamma$
modes, while the connection just sketched of this model with the HLS 
approach provides the lacking algebraic connection.

\section{SU(3) Breaking of the HLS--FKTUY Model}
\label{HLSmodel}  

\indent \indent The SU(3) symmetry breaking (referred to as BKY) we use
originates from Refs. \cite{HLS,BKY}. Briefs accounts and some 
new developments can be found in Refs.~\cite{BGP,heath},
connected more precisely with the anomalous sector \cite{FKTUY}.

\subsection{SU(3) Breaking Mechanism of the HLS Model}

\indent \indent 
Basically, the SU(3) breaking scheme we use has been introduced 
by Bando, Kugo and Yamawaki \cite{BKY} (referred to as BKY) 
and has given rise to a few variants \cite{BGP,heath}.
In order  to recover the charge normalization$F_{K^+}(0)=1$, even after breaking of SU(3)
flavor symmetry, one is obliged to define
a renormalized  pseudoscalar field matrix $P'$ in terms
of the bare one $P$ by
\begin{equation} 
P'=X_A^{1/2} P X_A^{1/2},
\label{brk1}
\end{equation}
where the breaking matrix $X_A$ writes diag($1,~1,~1+c_A$)
and we have \cite{BKY,heath}
\begin{equation} 
\ell_A \equiv  1+c_A=\left(\frac{f_K}{f_{\pi}}\right)^2= 
1.495 \pm 0.030 ~~~,
\label{brk2}
\end{equation}

The numerical value just given is deduced from 
the experimental information quoted in Ref.~\cite{PDG98}.

\subsection{A Phenomenological Lagrangian for Radiative Decays}
\label{anomalous}

\indent \indent Following FKTUY \cite{FKTUY},
the anomalous  U(3) symmetric Lagrangian describing
$PVV$ interactions and, using the non--anomalous Lagrangian \cite{heath,rad},
 $P V \gamma$ and $P \gamma \gamma$ transitions is given by Eq. (\ref{fktuy0}).
 It full expansion can be found in Ref. \cite{heath}.
Postulating that the same formulae apply when breaking nonet symmetry ($x \ne 1$),
is confirmed by its formal agreement with the O'Donnell derivation
of the coupling constant formulae.  
 
However, breaking the SU(3) symmetry \`a la BKY, also implies that
we have to reexpress the FKTUY Lagrangian in terms of the renormalized
matrix $P'$, instead of the bare one $P$; this is done using Rel. (\ref{brk1}),
with the (fixed) parameter given in Rel. (\ref{brk2}). We remind that
breaking nonet symmetry means that
$P'$ is also modified by the replacement $\eta_0 \longrightarrow x \eta_0 $. 
 
Propagating this field renormalization down to the FKTUY Lagrangian  writes
\begin{equation}
{\cal L}=- \frac{3 g^2}{4 \pi^2 f_{\pi}} 
\epsilon^{\mu \nu \rho \sigma} {\rm Tr}[\pa_\mu V_\nu\pa_\rho V_\sigma X_A^{-1/2}P'X_A^{-1/2}].
\label{wz2}
\end{equation}

Then, the VVP Lagrangian is changed in a definite way by the symmetry
breaking parameter $\ell_A$ defined above (see Eq.~(\ref{brk2})) and 
supposed to have a well understood numerical value 
(practically 1.5). 

The expanded form of this Lagrangian is given in the Appendix of Ref. \cite{rad}.
In principle, from this Lagrangian,  one is able to construct 
the decay amplitudes for the 
$V \rightarrow P \gamma$, $P \rightarrow V \gamma$, $V \rightarrow e^+e^-$
and $P \rightarrow  \gamma \gamma$ processes. They can be found in the
appendix in Ref. \cite{rad}.

\subsection{The VMD Description of $\eta/\eta' \rightarrow \gamma \gamma$ Decays}

\indent \indent
The expression for the decay amplitude 
 $G_{\eta\gamma \gamma}$ are given by Eq (42) in Ref. \cite{rad}.
It compares well with the corresponding expression of Ref.~\cite{takizawa}
deduced from the Nambu--Jona--Lasinio model. This shows that  breaking
parameters in this reference, originally expressed as 
functions of effective quark masses,  also get an  expression
in terms of $f_{\pi}/f_K$. More precisely, as remarked in Ref.  
\cite{takizawa}, in the (chiral) limit of vanishing 
meson masses, their breaking parameter, which can be  formally
identified to our $Z=[f_{\pi}/f_K]^2$, is simply the ratio $m_q/m_s$
($q$ stands for either of $u$ or $d$ which have equal masses if
SU(2) flavor symmetry is fulfilled)
of the relevant effective masses of quarks.

With this respect, a surprising connection could be made with the
traditional description of radiative decays using
quark magnetic moments \cite{dolinsky}. Indeed,
the present fit values for these are \cite{volodia}~:

\begin{equation}   
\begin{array}{llll}
\mu_u=1.852 & \mu_d=-0.972 & \mu_s=-0.630
\end{array}
\end{equation}

\noindent in units of Bohr magnetons. These magnetic
moments corresponds to the following quark (effective) masses

\begin{equation}   
\begin{array}{llll}
m_u=355.1~ {\rm MeV} & m_d=337.4 ~ {\rm MeV}& m_s=522.8~ {\rm MeV}
\end{array}
\end{equation}

It is indeed a point that $m_s/m_u=1.47$,
$m_s/m_d=1.55$ compare well with $[f_K/f_{\pi}]^2=1.495$,
as it can be guessed from the remark by Takizawa {\it et al.}
\cite{takizawa}. Whether, this is accidental, or
reveals a deeper property is an open question.

\subsection{The WZW Description of $\eta/\eta' \rightarrow \gamma \gamma$ Decays}
\label{wzwl1}

\indent \indent More interesting is that, starting from 
broken HLS and FKTUY, 
we recover the traditional form for these amplitudes,
({\it i.e.} the one mixing angle expressions of
Current Algebra \cite{DHL,GILMAN,chan}).
Using these standard expressions, one indeed gets  
through identification~:
\begin{equation}
\begin{array}{ll}
\displaystyle \frac{f_{\pi}}{f_8}=\frac{5-2Z}{3}~~, & ~~~
\displaystyle \frac{f_{\pi}}{f_0}=\frac{5+Z}{6}x~~~,
\end{array}
\label{wz6}
\end{equation}
where $Z=[f_{\pi}/f_K]^2$.
This shows that, in the limit of SU(3) symmetry, we have $f_8=f_{\pi}$ and
$f_0=f_{\pi}/x$, and that $f_0=f_8=f_{\pi}$ supposes  
that there is no symmetry breaking at all.

Actually, these formulae mean that, instead of going through
the whole machinery of VMD by starting from the broken
FKTUY Lagrangian, one could get these
coupling constants for two--photon decays 
of the pseudoscalar mesons by starting from the 
WZW Lagrangian \cite{WZW1,WZW2}. Indeed, this can be written

\begin{equation}
{\cal L}_{WZW} = -\frac{e^2}{4 \pi^2 f_{\pi}} \epsilon^{\mu \nu \rho \sigma}
\partial_{\mu}A_{\nu} \partial_{\rho}A_{\sigma} {\rm Tr} [Q^2P]
\label{wzw}
\end{equation}

\noindent where $Q=$ diag(2/3,--1/3,--1/3)  is the quark charge matrix,
$A$ the electromagnetic field
and $P$  is the bare pseudoscalar field matrix. Changing to $P'$ through
Rel. (\ref{brk1}), allows indeed to recover directly (and trivially)
the couplings given in the Appendix.

This illustrates clearly that, what is named $f_8$
in the Current Algebra \cite{chan} expressions for
$\eta/\eta'$ decays to two photons, can be expressed solely in 
terms of $f_{\pi}$
and $f_K$, in a way which fixes its value to $f_8=0.82 f_{\pi}$.
The fact that the WZW Lagrangian leads to the same results as the FKTUY 
Lagrangian simply states their expected equivalence when deriving two--photon 
decays amplitudes for pseudoscalar mesons.

On the other hand, the SU(3) sector of Chiral Perturbation Theory (ChPT)
predicts \cite{DHL,GL85} $f_8/f_{\pi} \simeq 1.25$. One could thus
think  to a contradiction \cite{rad} between VMD (or FKTUY) and WZW
on the one hand, and ChPT on the other hand. However, it happens \cite{chpt}
that this is a misleading appearence due to an inconsistency between
defining the decay constants and mixing angle in agreement with
Current Algebra -- definitions recovered by VMD and WZW as illustrated
above -- and current ChPT definitions.

\section{Fitting Decays Modes with the Broken Model}
\label{fitres}
\indent \indent
In this section, we focus on the model for coupling constants
given by  Eqs.~(39) to (45)  of Ref \cite{rad}, and  use them
for a fit to radiative decays of light mesons. the corresponding data
are all taken{\footnote{We have however used in the fits \cite{rad},
as partial width for $\eta \rightarrow \gamma \gamma$, the mean value
of the measurements reported by $\gamma \gamma$ experiments, instead 
of the PDG \cite{PDG98} value, which is affected by the single existing 
Primakoff measurement.}}
from the Review of Particle Properties \cite{PDG98}.

 From what is reported several times in the literature 
\cite{volodia,ben2,ball}, one might expect
potential problems with one or both $K^*$ decay modes. 
Therefore, we have followed the strategy of performing fits
of all radiative decay modes except for these two. 
In all fits performed with the model described above, we have found 
that the prediction for $K^{*0} \rightarrow K^0 \gamma$ is in fairly good 
agreement with the corresponding measurement \cite{PDG98}, while the 
expected value for $K^{*+} \rightarrow K^+ \gamma$ is always
at about $5\sigma$ from the accepted value \cite{PDG98}. Therefore,
in the fits referred to hereafter, the process $K^{*+} \rightarrow K^+ \gamma$
has been removed. 

The difficulty met with this decay mode in several studies mentioned above
could cast some doubt on the reliability of this measurement, performed
using the Primakoff effect. However, it cannot be excluded that this measurement 
is indeed correct and that the reported disagreement simply points toward the need of 
refining the  models.  This will be discussed in Section \ref{kstar}.

\subsection{SU(3) Breaking and the Value of $f_K$}

\indent \indent The key parameter associated specifically with  the breaking
of  SU(3) flavor symmetry is the BKY parameter  $\ell_A$, expected to 
be equal to $[f_K/f_{\pi}]^2$ (see Rel. (\ref{brk2})) . As starting point in our 
fit, we have left free all parameters~: $G$, $x$, $\theta_V$, $\theta_P$ 
and $\ell_A$. We thus got a nice fit probability ($\chi^2/$dof$=10.74/9$) 
and the result we like to mention from this fit is

\begin{equation}
\displaystyle \frac{f_K}{f_{\pi}}=1.217^{+0.021}_{-0.019}. 
\label{fkfpi}
\end{equation}
which is almost exactly the value expected from the known ratio $f_K/f_{\pi}$.
This gives, of course, a strong support to the BKY breaking
mechanism  \cite{BKY,heath}. 

This result strongly suggests that one can reasonably  
{\em fix}  $\ell_A=1.50$ (at its physical value). Then, the single free
breaking parameter which influences the coupling constants in radiative
decays, beside mixing angles, is the nonet symmetry breaking parameter $x$.

\subsection{The Fit Parameter Values}

\indent \indent
Except for
the two mixing angles, we only have two free parameters to fit the data set,
as in the unbroken case \cite{ben2}.
One,  $G$, is connected with the vector meson universal coupling $g$, the other
is the nonet symmetry breaking  parameter $x$. 
The final fit exhibits a very good quality ($\chi^2/dof=10.9/10$),
corresponding to a $44\%$ probability, and the
best values and errors for the main parameters ($\ell_A$  is
fixed to 1.5) are

\begin{equation}
\left\{
\begin{array}{llll}
G &= 0.704 \pm 0.002 & [{\rm GeV}]^{-1}\\[0.5cm]
x &= 0.917 \pm 0.017  &~~\\[0.5cm]
\theta_V &= ~~31.92 \pm 0.17 &[{\rm deg.}]\\[0.5cm]
\theta_P &= -11.59 \pm 0.76 &[{\rm deg.}]
\end{array}
\right.
\label{fitx}
\end{equation}

The nonet symmetry breaking parameter is $x=0.92 \pm 0.02$,
confirming a previous analysis \cite{ben2}.  The value for $G$
is also in nice agreement with the previous analysis of 
Ref. \cite{ben2}. The vector mixing angle
is found at 3.4 degrees below its ideal value{\footnote{
Let us, however, remind  that this value relative to ideal mixing
is the consequence of our choice $\phi^I=-|s\overline{s}>$.}}
and agrees with predictions \cite{DM}.
The mixing angle of pseudoscalar mesons coming out from fit 
points toward a small deviation from the Gell--Mann--Okubo mass relation
\cite{DHL,chpt}.

\subsection{The One Angle $\eta/\eta'$ Mixing Scheme from VMD}
\label{mixing}

\indent \indent
As discussed above the model we propose, which 
relies on the VMD approach of 
Refs.~\cite{HLS,FKTUY} with fixed SU(3) breaking
\`a la BKY \cite{BKY,heath}, leads to  (one angle)
formulae for the $\eta/\eta' \rightarrow \gamma \gamma$ decay amplitudes.
These can be identified with the corresponding Current Algebra
standard expressions and we have recalled that they can also be directly
derived from the WZW Lagrangian. This justifies the identification shown in 
Eq.~(\ref{wz6}) for the singlet and octet coupling constants.
One should note that nonet symmetry breaking does not 
modify the formulae substantially.
In this case, we obtain together with  
$\theta_P=-11.59^{\circ} \pm 0.76^{\circ}$~:
\begin{equation}
\begin{array}{ll}
\displaystyle \frac{f_8}{f_{\pi}}=0.82 \pm 0.02~~~, & 
~~  \displaystyle \frac{f_1}{f_{\pi}}=1.15 \pm 0.02
\end{array}
\label{f1f8}
\end{equation}
using Eq.~(\ref{fkfpi}), and the fit result for $x$.

Actually, it happens that a low value for $f_8/f_{\pi}$
and a low absolute value for the pseudoscalar mixing angle $\theta_P$
are correlated properties\cite{ball,rad}.

One can also perform a fit of
the $VP\gamma$ processes in isolation in order to get
estimates for $x$ and $\theta_P$, free of any influence
of the  $P \gamma \gamma$ processes. This allows to
check the conceptual relation between $VP\gamma$
and $P \gamma \gamma$  inferred from VMD.
Using the formulae of Ref. \cite{rad}, 
one can indeed reconstruct the VMD expectations
for the $P \gamma \gamma$ modes. The interesting  point here, compared
with what is shown in Table 1, is  that the fit procedure
improves the parameter values associated with  the $VP\gamma$ modes.
The results are shown in Table 2.

Then, the HLS approach we have developed, even restricted to
the $VP\gamma$ processes is indeed able to predict
quite nicely the  $P \gamma \gamma$ partial widths. The effects
of fitting can easily be understood by comparing the corresponding
information and accuracies in Tables 1 and 2. Thus, the VMD formulae (which are also 
those obtained \cite{heath} by breaking \`a la BKY the Wess-Zumino Lagrangian)
provide  the traditional one--angle scheme of the former Current Algebra. 
No need for a more complicated mixing pattern  \cite{leutw,leutwb} emerges 
from the data on radiative decays.

\begin{center} 
\begin{tabular}{|| c  | c  | c | c | c | c ||}
\hline
\hline
\hhhc Mode & VMD Fit  &  PDG & Comment\\
\hline
\hline
 \hhhd~~~&  & $0.514 \pm 0.026$ & $\gamma \gamma$ \\[0.5cm]
$\eta \rightarrow \gamma \gamma$ [keV]&
$0.464 \pm 0.026$ & $0.46 \pm 0.04$ & PDG mean \\[0.5cm]
 ~~~&  & $0.324 \pm 0.046$ & Primakoff \\[0.5cm] 
\hline
$\eta' \rightarrow \gamma \gamma$\hhhu[keV] &
$4.407 \pm 0.233$ & $4.27 \pm 0.19$ & PDG mean\\ 
\hline
\end{tabular}
 
\parbox[t]{16.0cm}{ \hhht
      {\bf Table 2} :  Partial decay widths of the $\eta/\eta'$
      mesons, as reconstructed from fit to solely the radiative decays
      $VP\gamma$ (leftmost data column) and  their direct  measurements
      \cite{PDG98} (rightmost data column). 
}  
\end{center}

\section{Is There a Glue Component Coupled to $\eta/\eta'$?}
\label{nsglue}

\indent \indent
As stated in the Introduction, the precise content of the pseudoscalar singlet
component in $\eta/\eta'$ mesons is  somehow controversial. One cannot indeed 
exclude the interplay of the usual singlet 
$\eta_0=(u \overline{u}+d \overline{d}+ s \overline{s})/\sqrt{3}$
with other SU(3) singlet states \cite{ball,veneziano}, which could be
glueballs or some $c\overline{c}$ admixture, or both. Let us assume the existence
of such an additional singlet state that will be denoted  $gg$, in order
to make formally the connection with its possibly being a gluonium.
Then we should allow for mixing of these two possible singlet component
with the standard SU(3) octet 
$\pi_8=(u \overline{u}+d \overline{d}-2 s \overline{s})/\sqrt{6}$

\subsection{The $\eta/\eta'$ Mesons in Terms of Octet and Singlet States}

\indent \indent

An appropriate parametrization for the mixing of $(\pi_8, \eta_0, gg)$ into 
physical pseudoscalar meson states denoted $(\eta, \eta', \eta'')$ is needed.
The symbol $\eta''$ for the third partner
of the doublet $(\eta, \eta')$ simply means that we consider premature to try identifying it.
This is  done by means of an orthogonal matrix transform~; an  
appropriate parametrization of this matrix is 
the Cabibbo--Kobayashi--Maskawa matrix (with  no complex phase). 
 This tranform is given by Eq (2) in Ref. cite{rad}.
It depends on 2 angles, in addition to the usual $\theta_P$, which have been
named $\beta$ and $\gamma$. They are such that
the vanishing of  $\beta$ and $\gamma$ gives smoothly the usual mixing
pattern of the $(\eta,\eta')$ doublet (with one angle $\theta_P$)
and the decoupling of the additional singlet. Setting $\beta=0$
cancels out glue inside $\eta$ only, while $\gamma=0$ removes any glue
inside the $\eta'$ only. Then this transform 
allows for analyzing the interplay of an additional singlet
(named here glue) in a continuous way for both the $\eta$
and the $\eta'$ mesons.

\subsection{Nonet Symmetry Breaking versus Glue}

\indent \indent Up to now, we have illustrated that
the BKY breaking was a fundamental tool in order to describe
all data concerning radiative and two--photon decays of light
mesons. The other central result of our fitting model concerns the
unavoidable need of about 10\% breaking of nonet symmetry 
in the pseudoscalar sector ($x \simeq 0.9$). 
This could well be a fundamental property. 

However, the observed nonet symmetry breaking could also be an 
artefact of the model above, reflecting physical effects intrinsically 
ignored. In this Section we examine the interplay of nonet
symmetry breaking and a possible glue component.
 The $VP\gamma$ and $P\gamma \gamma$ coupling constants have been determined  
(see Eqs (46) and (47) in Ref. \cite{rad}).

A  phenomenological study of these relations, which includes  SU(3)
breaking, nonet symmetry breaking and glue has been performed with 
the following conclusions~: {\bf 1} The BKY breaking is still found
determined by the value of $f_K/f_{\pi}$~; it can thus be fixed as previously done.
 {\bf 2} Nonet symmetry breaking 
and glue are intimately connected and reveal a correlation close to the 100\% level.
This second remark does not mean that nonet symmetry breaking 
and glue (or any additional singlet) are physically equivalent.
The single appropriate conclusion is rather that, in order to conclude 
firmly about each  of these twin phenomena, one needs relatively
precise information on the other \cite{chpt}.  

\begin{figure}[tbp]
  \centering{\
     \epsfig{angle=0,figure=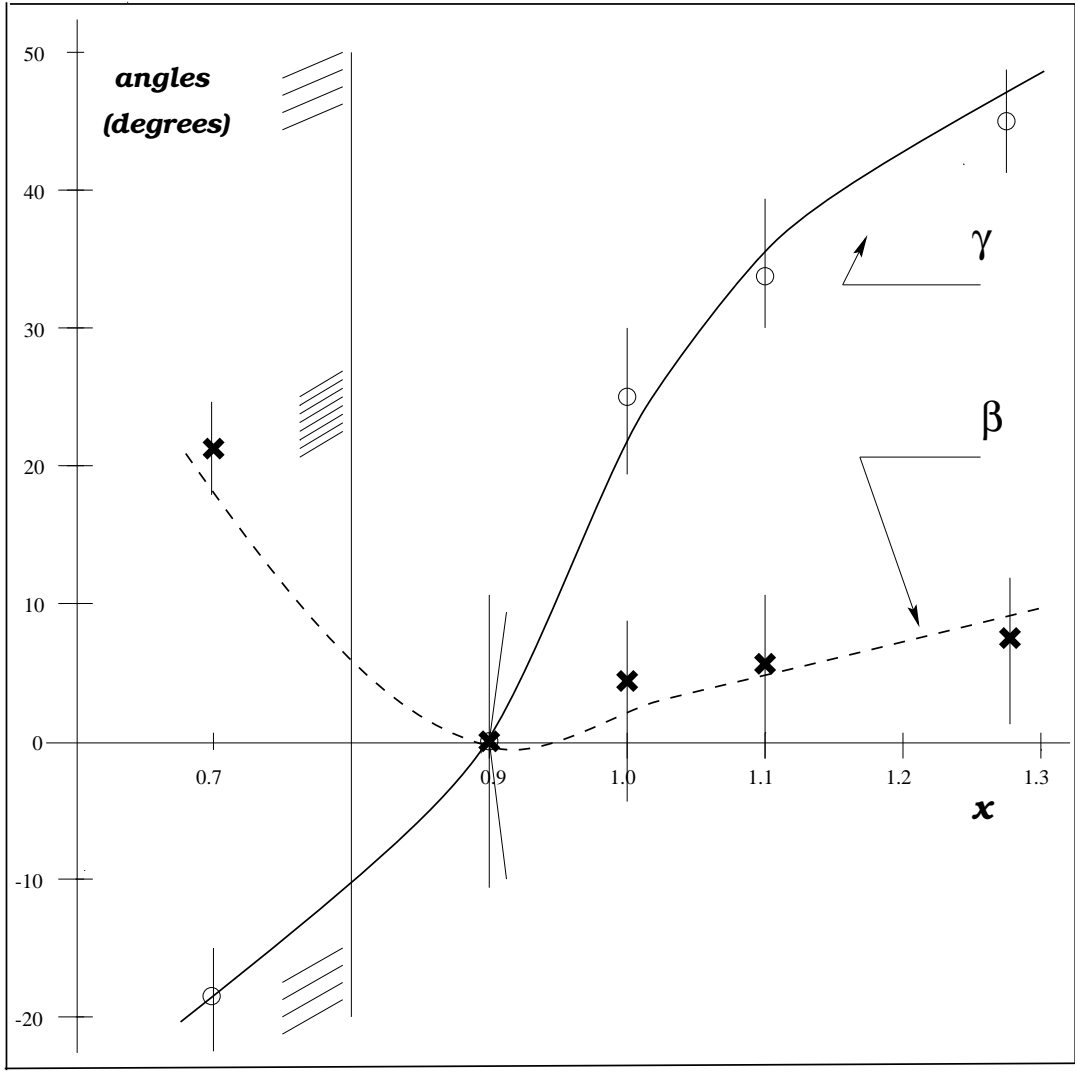,width=0.9\linewidth}
               }
\parbox{130mm}{\caption
{The angles providing the coupling of $\eta$ and $\eta'$ to
glue (actually, to any singlet state not constituted of 
$u$, $d$ and $s$ quarks) as a function of the symmetry breaking parameter $x$.
$x=1$ corresponds to exact nonet symmetry for the pseudoscalar mesons
couplings. A non--zero $\beta$ is tightly connected with glue in the $\eta$
meson, while a non--zero $\gamma$ is tightly connected with glue in the $\eta'$
meson.}
\label{angle}}
\end{figure}

However, a few additional remarks can be drawn \cite{rad}.
One can analyze how coupling to glue evolves as a function of
a fixed nonet symmetry breaking level. This is shown in Fig \ref{angle}.
One clearly sees there that no need for glue is exhibited 
by the data if $x \simeq 0.9$ ($\beta$ and $\gamma$ can be
chosen equal zero without hurting the data). 
At $x \simeq 1$ and somewhat above, the angle
$\beta$ is still consistent with zero, pointing to the fact
that one can hardly claim the need for glue in the $\eta$ meson.
However, somewhat above $x \simeq 0.9$, the level of glue in $\eta'$
is a rising function of $x$, as shown by the steep dependence of
$\gamma$ upon $x$. In order to fix one's idea,
if for some reason $x \simeq 1$ has to be prefered, then 
one can express the glue
fraction in $\eta'$ by $\cos^2{\gamma}\simeq 0.20$ (at $x=1$).

In view of all this, beside the model with no glue and
with a small breaking of nonet symmetry, we have studied
the case of glue in only the $\eta'$ meson (setting $\beta=0$)
and no breaking of nonet symmetry. Moreover, the above remarks  
justify to perform a fit by fixing (as before) the SU(3) 
breaking at its expected value ($\ell_A=1.5$), choosing also 
$x=1$ and  $\beta=0$ (in order to lessen at most correlation effects).
In this case we have exactly the same number of parameters as
in the previous set of fits.  The corresponding fit results
show a nice quality ($\chi^2/dof=10.5/10$), quite
equivalent to the no--glue case. The predicted branching fractions are
discussed in Section \ref{prediction} below.

As major conclusions of this section, one can first assert that a possible glue 
content inside the $\eta$ is not requested by the data. A significant
glue content inside the $\eta'$ is possible, however subject to the actual level
of nonet symmetry breaking \cite{chpt}. 

\section{Estimates for Branching Fractions from Fits}
\label{prediction}

\indent \indent  We give and discuss here the reconstruction
properties of the the two variants of our model, both discussed above.
 These are  {\bf i/}
nonet symmetry breaking supplemented  by a fixed SU(3) breaking (BKY) 
and  {\bf ii/} a fixed SU(3) breaking (BKY) 
with glue inside the $\eta'$ replacing nonet symmetry breaking.

We now compare the branching fractions predicted by these two 
solutions to the accepted branching fractions as given in the
Review of Particle Properties \cite{PDG98}. They are computed
according to the formulae for coupling constants given in Ref. \cite{rad}.
In Table {\bf 3}, we list the information for radiative decays.
The first remark which comes to mind by comparing the two model reconstructions
is that their predictions are close together (see the first two data column).
This illustrates clearly the numerical equivalence of coupling to glue and
nonet symmetry breaking.

The relative  disagreement of  $\eta' \rightarrow \rho^0 \gamma$
with accepted values \cite{PDG98} is actually an interesting artefact. 
Indeed, what has been submitted to fit is not 
the branching fraction given in Ref. \cite{PDG98}, but
the corresponding coupling constant  extracted
by the Crystal Barrel Collaboration in \cite{abele}. The reason
for this is that the (published) branching fraction for 
$\eta' \rightarrow \rho^0 \gamma$  is influenced by a non--resonant
contribution originating from the box anomaly
\cite{ben2,chan,WZW1,WZW2} for the vertex
$\eta' \pi^+ \pi^- \gamma$.  This is not accounted for in the VMD model
of \cite{FKTUY} and has thus to be removed. 
Actually this process contributes to the total $\chi^2$
by only $\simeq 0.5$. Moreover, its importance is not that decisive
that it influences the fit results dramatically. This last information,
has been tested by removing the $\eta' \rightarrow \rho^0 \gamma$
decay from fit data. 

On the other hand, even if quite acceptable, the reconstruction
for $\eta \ra \gamma \gamma$ branching fraction is influenced
by having used for this decay mode the mean value of the measurements
obtained in $\gamma \gamma$ experiments, while
the PDG information reported for $\eta \ra \gamma \gamma$
branching fraction is the official one \cite{PDG98}, somehow influenced
by the Primakoff measurement. 

The single clear disagreement of model predictions with  data  
concerns the branching fraction for $K^{*\pm} \rightarrow K^{\pm} \gamma$,
that we find about half of the reported value
in the Review of Particle Properties \cite{PDG98}.
We postpone to Section \ref{kstar} the reexamination of this question. 

The recent measurements for $\phi \rightarrow \eta' \gamma$
are also well accepted by the fit. However, the prediction
tends to indicate that the central value found by SND Collaboration
\cite{phietp2} is favored compared to that of the
CMD2 Collaboration \cite{phietp1}.

All this leads us to conclude that the model of symmetry
breaking we have presented provides a consistent description
of the data. At their present level of accuracy, these
do not seem to require additional symmetry breaking effects.
An especially satisfactory conclusion is that  SU(3)
breaking effects are not left free in the fits and are
practically determined by the ratio $f_K/f_{\pi}$.
Some nonet symmetry breaking and/or glue is needed
(see however Ref. \cite{chpt}).
 
\section{The $K^{*\pm}$ Radiative Decay Problem}
\label{kstar}

\indent \indent As shown by the two leftmost data columns in
Table {\bf 3}, the two variants of the model presented above
do not account for the accepted \cite{PDG98}
radiative decay width  $K^{*\pm} \rightarrow K^{\pm} \gamma$.
One cannot exclude that this measurement might have to
be improved by other means that the Primakoff effect.
However, this decay mode has been measured separately 
for the two charged modes and found to agree with each
other. Therefore, the possibility that this failure
indicates that models have to be refined cannot be avoided.

The first point which comes to mind is whether the disagreement
reported above (a factor of two between prediction and measurement) 
could be attributed to (missing) SU(2) flavor symmetry breaking effects.
If one takes into account the quark content of the $K^*$'s,
the answer is seemingly no. Indeed, in this case, one could 
guess that significant unaccounted for SU(2) breaking effects would
have rather affected the quality of predictions for $K^{*0}$ rather
than for $K^{*\pm}$. 
  
This possibility seeming unlikely, the question becomes~:
can the VMD modelling we developed  be modified
in order to account for this mode within an extended SU(3) breaking 
framework? The reply is positive and is the following. 

\subsection{The $K^*$ Model}

\indent \indent
Within the  spirit of the BKY mechanism, the (unbroken)
FKTUY Lagrangian given in Rel. (\ref{fktuy0}) can be broken
straightforwardly in three different ways. The first
mean is the pseudoscalar field renormalization
(see Rel. (\ref{wz2})), which leads to introduce the 
matrix $X_A$ and thus the breaking parameter
$\ell_A$ found equal to $(f_K/f_{\pi})^2$ as expected 
\cite{BKY}. It has been successfully supplemented with nonet 
symmetry breaking (and/or glue). 

A second mean has been proposed  by Ref.
\cite{BGP} (referred to as BGP breaking).
It turns to introduce a breaking matrix $X_W=$ diag(1, 1, $1+c_W$) 
and a new breaking parameter $\ell_W=1+c_W$ in a symmetric way
inside the FKTUY Lagrangian{\footnote{The symmetry can be made
manifest. What is written in Rel. (\ref{bgp}), can be symbolically
written Tr$[VX_WV(X_A^{-1/2}PX_A^{-1/2})]$ and is obviously identical
to  Tr$[X_W^{1/2}V(X_A^{-1/2}PX_A^{-1/2})VX_W^{1/2}]$.}}~:
 
\begin{equation}
{\cal L}=- \frac{3 g^2}{4 \pi^2 f_{\pi}} 
\epsilon^{\mu \nu \rho \sigma} {\rm Tr}[\pa_\mu V_\nu X_W\pa_\rho 
V_\sigma X_A^{-1/2}P'X_A^{-1/2}]
\label{bgp}
\end{equation}

In Ref. \cite{rad}, it has been shown that, supplementing
the BKY breaking $X_A$, the BGP breaking $X_W$ alone is unable 
to account for the $K^{*\pm}$ radiative decays. Moreover,
the constant $\ell_W$ is pushed to 1 by the fit procedure,
and then to no BGP breaking  ($c_W=0$).  

A third mean is however conceivable.
One should note that the BKY breaking mechanism \cite{BKY} implies a renormalization
(or redefinition) of the pseudoscalar field matrix expressed through $X_A$~; however,
the $X_V$ breaking \cite{BKY,heath} does not end up with a renormalization of the vector 
field matrix, which remains unchanged in the breaking procedure. One can then
{\it postulate} that the vector meson field matrix has also to be SU(3) broken and
also in a symmetric way. This is done by performing the change~:

\begin{equation}
V \longrightarrow X_T V X_T~~,~~~[~X_T=\rm{diag}(1,1,1+c_T)~]
\label{wzp1}
\end{equation}

\noindent in Rel. (\ref{bgp}), in complete analogy with the renormalization
of the $P$ matrix. A lack of fancy (not still a mathematical proof)
seems to indicate that no fourth mechanism can play.

A detailed study of the consequences of Lagrangian (\ref{wzp1}) has been
performed in Ref. \cite{rad} with an interesting conclusion~: if one fixes
$\ell_A=1.5$ as expected \cite{BKY,heath}, the $X_T$ and $X_W$ breaking
are so sharply correlated that they cannot be left free together. More
interestingly, it was found phenomenologically that the additional
breaking parameters fulfill~:

\begin{equation}
(1+c_W) (1+c_T)^4=1 
\label{wzp4}
\end{equation}

This tells us that the most general form of the broken FKTUY Lagrangian 
accepted by the data can be symbolically written~:
 
\begin{equation}
{\cal L}=C ~{\rm Tr}[(X_T^{-1} V X_T)(X_A^{-1/2}P'X_A^{-1/2})(X_T V X_T^{-1})]
\label{wzbrka}
\end{equation}

In this case, all couplings constants write as when having solely the BKY breaking
mechanism, supplemented with nonet symmetry breaking and/or glue, except for the $K^*$
decay modes which become~: 

\begin{equation}
\left \{
\begin{array}{lll}
G_{K^{*0} K^0 \gamma}=&- & \displaystyle G \frac{\sqrt{K'}}{3} (1+\frac{1}{\ell_T} )~ \\[0.3cm]
G_{K^{*\pm} K^{\pm} \gamma}=& &\displaystyle G \frac{\sqrt{K'}}{3}(2-\frac{1}{\ell_T})  
\end{array}
\right.
\label{wzp2}
\end{equation}

\noindent where $K'=\ell_T/\ell_A$ and $\ell_T=(1+c_T)^2$.
Stated otherwise, both $K^*$ couplings are changed~: the one correctly
accounted for by the previous modellings ($K^{*0}$), {\it and} the
one poorly described ($K^{*\pm}$). Therefore, a fit
value for $\ell_T$ must change $G_{K^{*\pm}}$ while leaving
$G_{K^{*0}}$ practically unchanged, despite the {\it functional} relation
among them.

Assuming no coupling to glue, we have performed the fit and found
a perfect fit quality ($\chi^2/dof=11.07/10$) with practically
the same parameter values as in the models above and additionally~:

\begin{equation}
\ell_T = 1.19 \pm 0.06~~~, ~~(c_T=0.109 \pm 0.024)
\label{wzp5}
\end{equation}

The predicted branching fractions are given in the third data column in Table {\bf 3}.
They indeed show that all predictions (including for the $K^{*0}$ mode) are unaffected
except for the $K^{*\pm}$ mode, now in quite nice agreement with its accepted value 
\cite{PDG98}.

One may wonder that the $K^{*0}$ mode is unchanged, while the $K^{*\pm}$ mode
is increased by a factor $\simeq 2$. For this purpose,
one may compare the values of the $\ell_T$
part of the couplings in Rels. (\ref{wzp2}) at $\ell_T=1$ and at $\ell_T=1.2$.
One thus find that the former change is  $ 2 \rightarrow 2.01$, while the later
change is $1 \rightarrow 1.28$. Therefore, the change requested in order
to account for the $K^{*\pm}$ mode, results in an unsignificant change for
the $K^{*0}$ mode.

Therefore, quite unexpectedly, a tiny change in the VMD model we
have shown is enough to describe indeed all radiative decays at 
their presently accepted values.

Nevertheless, the additional mechanism complicates the full breaking picture
which is otherwise quite simple. One can hope that new measurements
for the $K^{*\pm}$ radiative decay will come soon and tell definetely
whether this complication really proceeds from physics.

\subsection{The $K^*$ Model and the WZW Lagrangian}

\indent \indent In Section \ref{wzwl1}, we have remarked that
imposing the change of fields from $P$ to $P'$ to the WZW
Lagrangian (see Rel. (\ref{wzw})) provides the same description of 
radiative decays of pseudoscalar mesons than the broken HLS--FKTUY
model. Thus, one has checked that that these two descriptions
were indeed equivalent. The VMD description is however able to
connect the $P\gamma \gamma$ couplings to the $VP\gamma$ ones
with the success illustrated by Table {\bf 2}.

When introducing the additional breaking schemes in order
to construct the $K^*$ model sketched above (the expanded
Lagrangian can be found in the Appendix of Ref. \cite{rad}),
this property is formally lost, except if additionally
to the change $P \rightarrow P'$, we also perform the change
$Q^2 \rightarrow X_WX_T^4 Q^2$, e.g. if we ``renormalize''
the SU(3) charge matrix, or the WZW Lagrangian as a whole.
It happpens that the condition in Rel. (\ref{wzp4})
prevents such an ugly transform. Stated otherwise,
phenomenology forces a relation which is such that the two--photon
decays of pseudoscalar mesons are still given by the
Wess--Zumino--Witten Lagrangian \cite{WZW1,WZW2}, with
breaking only for the single occuring matter field matrix $P$.

In order  that the equivalence between VMD and WZW
is generally maintained, the $K^*$ breaking should
affect the  $K^*$ couplings only. If, instead, the
other couplings $VP\gamma$ were affected, this would
propagate down to the $P\gamma\gamma$ couplings.
In this case, the equivalence statement between
(broken) VMD and (broken) WZW would no longer hold.
Thus, the $K^*$ breaking scheme seems indeed 
the most general consistent with this equivalence statement.

\section{Conclusion}
\label{conclud}

\indent \indent
The VMD based model we have presented indeed
describes all radiative decays $VP\gamma$ and $P \gamma\gamma$
accurately. 
This model relies  heavily on the HLS model supplemented with
the BKY breaking mechanism in order to account for SU(3) symmetry breaking.
 It should be stressed that
phenomenology indeed confirms the theoretical connection
between the SU(3) breaking parameter and the ratio 
$f_K/f_{\pi}$.
 
One has additionally to introduce a further degree of
freedom~; this can be either of direct nonet symmetry breaking
($x \ne1 $) or coupling of the $\eta'$ meson to an additional
singlet ($\gamma \ne 0$). A mixture of both effects
is also an acceptable solution, numerically and theoretically. 
Actually, radiative decays of light mesons alone cannot
provide more detailed information about this possible additional singlet 
without additional theoretical input.

The picture that emerges from there is quite consistent
and tends to indicate that present data do not require
any breaking of the SU(2) symmetry at a visible level
in only radiative decays of light mesons.

The single present datum which requires special additional input
is  the $K^{*\pm}$ radiative decay. It can be done succesfully
without destroying an equivalence statement between
VMD and the WZW description of pseudoscalar meson
decays to two photons. However, a confirmation of
the present data for the $K^{*\pm}$ radiative decay
looks desirable.

Anyway, whatever is the precise value of the
$K^{*\pm}$ radiative width, VMD expressed through
the general concept underlying the HLS model
is able to provide
a quite consistent picture of the radiative decays
of all light mesons.

\vspace{1.0cm}
\begin{center}
{\bf Acknowledgements}
\end{center}
HOC was supported by the
US Department of Energy under contract
DE--AC03--76SF00515. SE was supported by
the Division des Affaires Internationales of IN2P3 and would like to
thank
the LPNHE Laboratory for its hospitality; VNI was supported by
the Direction des Affaires Internationales of CNRS.
Both SE and VNI are grateful to
Eliane Perret (IN2P3) and
Marcel Banner (LPNHE) for their help and support.

\newpage 
\begin{center} 
\begin{tabular}{|| c  | c  | c | c | c ||}
\hline
\hline
\hhhc  \hhhb  Process      & Nonet Sym. & Glue      & $K^{*\pm}$ &  PDG \\
\hhhc  \hhhb               & $+$ SU(3)  & $+$ SU(3) & Breaking   &  \\
\hline
\hline
$\rho \rightarrow \pi^0 \gamma$  $(\times 10^4)$ &\hhhb $5.16 \pm 0.03$&$5.16 \pm 0.03$ & $5.16 \pm 0.03$
& $6.8 \pm 1.7$\\
\hline
$\rho \rightarrow \pi^\pm \gamma $  $(\times 10^4)$ & $5.12 \pm 0.03$ & $5.12 \pm 0.03$ & $5.12 \pm 0.03$
& $4.5 \pm 0.5$ \hhhu \\
\hline
\hline
$\rho \rightarrow  \eta \gamma $  $(\times 10^4)$ \hhhu & $3.25 \pm 0.10$ & $3.28 \pm 0.10$ &$3.31 \pm 0.09$
& $2.4^{+0.8}_{-0.9}$\\
\hline
$\eta' \rightarrow \rho \gamma$ $(\times 10^2)$\hhhu & $33.1 \pm 2.0$ & $33.7 \pm 2.0$ & $33.0 \pm 1.8$
& $30.2 \pm 1.3$ \\
\hline
\hline
$K^{*\pm} \rightarrow K^\pm \gamma $  $(\times 10^4)$ \hhhu  & $5.66 \pm 0.03$ & $5.66 \pm 0.03$ &$9.80 \pm 0.93$
& $9.9 \pm 0.9$ \\
\hline
$K^{*0} \rightarrow K^0 \gamma $  $(\times 10^3)$ \hhhu  & $2.30 \pm 0.01$ & $2.30 \pm 0.01$ &$2.32 \pm 0.02$
& $2.3 \pm 0.2$\\
\hline
\hline
$\omega \rightarrow \pi^0 \gamma$ $(\times 10^2)$\hhhu  & $8.50 \pm 0.05$ & $8.50 \pm 0.05$ & $8.50 \pm 0.05$
& $8.5 \pm 0.5$ \\
\hline
$\omega \rightarrow \eta \gamma$  $(\times 10^4)$ \hhhu  & $8.0 \pm 0.2$ & $8.1 \pm 0.2$ & $8.12 \pm 0.19$ 
& $6.5 \pm 1.0$\\
\hline
$\eta' \rightarrow \omega \gamma$ $(\times 10^2)$\hhhu  & $2.8 \pm 0.2$ & $2.9 \pm 0.2$ & $2.8 \pm 0.2$
& $3.01 \pm 0.30$ \\
\hline
\hline
$\phi \rightarrow \pi^0 \gamma $  $(\times 10^3)$ \hhhu & $1.27 \pm 0.13$ & $1.28 \pm 0.12$ & $1.26 \pm 0.13$ 
& $1.31 \pm 0.13$\\
\hline
$\phi \rightarrow \eta \gamma$ $(\times 10^2)$\hhhu  & $1.25 \pm 0.04$ & $1.25 \pm 0.05$ & $1.22 \pm 0.04$
& $1.26 \pm 0.06$ \\
\hline
$\phi \rightarrow \eta' \gamma $  $(\times 10^4)$ \hhhu  & $0.61 \pm 0.027$ & $0.55 \pm 0.03$ & $0.63 \pm 0.02$
&$1.2^{+0.7}_{-0.5}$\\
\hline
\hline
$\eta \rightarrow \gamma \gamma$ $(\times 10^2)$\hhhu  & $40.5 \pm 1.7$ & $40.8 \pm 1.8$ & $41.5 \pm 1.4$
& $39.21 \pm 0.34$ \\
\hline
$\eta' \rightarrow \gamma \gamma$ $(\times 10^2)$\hhhu  & $2.1 \pm 0.1$ & $2.1 \pm 0.1$ & $2.1 \pm 0.1$
& $2.11 \pm 0.13$ \\
\hline
\hline
\end{tabular}

\parbox[t]{16.0cm}{ \hhht
      {\bf Table 3} : Branching fractions from fits for radiative decays under various conditions of symmetry breakings.
Note that the rate for $K^{*\pm}$ is a prediction in the first two data columns,
while the corresponding data is included in the fit which  leads to the third data column.
}
\end{center}

\end{document}